\documentclass[aps,prb,twocolumn,showpacs,floatfix,superscriptaddress]{revtex4}
\usepackage{graphicx}
\usepackage{amssymb}
\usepackage{amsmath}
\usepackage{lmodern}
\usepackage{color}
\usepackage{hyperref}
\usepackage{simplewick}
\begin{document}

\preprint{This line only printed with preprint option}

\title{Replica symmetry breaking for  anisotropic  magnets with quenched disorder}

\author{E. Kogan}

\affiliation{Jack and Pearl Resnick Institute, Department of Physics, Bar-Ilan University, Ramat-Gan 52900, Israel}
\affiliation{Max-Planck-Institut fur Physik komplexer Systeme,  Dresden 01187, Germany}
\affiliation{Center for Theoretical Physics of Complex Systems, Institute for Basic Science (IBS), Daejeon 34051, Republic of Korea}

\author{M. Kaveh}

\affiliation{Jack and Pearl Resnick Institute, Department of Physics, Bar-Ilan University, Ramat-Gan 52900, Israel}
\affiliation{Cavendish Laboratory, University of Cambridge, J J Thomson Avenue, Cambridge CB3 0HE, UK}

\date{\today}

\begin{abstract}
We study critical behaviour of  a magnet  with  cubic anisotropy and quenched
scalar disorder which is taken into account by replica method.
We derive to first order in $\epsilon$ approximation the renormalization group  equations
taking into account possible replica symmetry breaking. We study the stability of the replica symmetric fixed points with respect to 
perturbations without (in general case) replica symmetry. However, we find that if a fixed point is stable with respect to replica 
symmetric deviations, it is also stable with respect to deviations without replica symmetry.
\end{abstract}

\pacs{64.60.ae,64.60.Ej,64.60.F-}

\maketitle

\section{Introduction}

Wilson and Fisher calculation of the critical exponents for the scalar $\phi^4$ (and $X-Y$)  model by analysis of the renormalization group (RG) equations \cite{wilson} was a groundbreaking discovery. The perturbation expansions of the beta functions of the RG equations turned out to be $\epsilon$ expansions,
and the rules how to calculate the coefficients in such expansions were formulated.
Immediately after that $O(n)$ model with cubic anisotropy term was studied by Aharony \cite{aharony} (see also Refs. \onlinecite{chaikin,izyumov,cardy}).
Later the $\epsilon$ expansion   was used
 to derive the RG equations for the vector $\phi^4$ model with   quenched scalar disorder \cite{khmel,lubensky,grinstein}  (see also  Ref. \onlinecite{ma}).
and  for   the model with cubic anisotropy  and such disorder,  also for the model with random direction of the anisotropy axis \cite{aharony2}.

The quenched disorder was taken into account in the mentioned above works by the replica  method \cite{dotsenko}. This method was advanced by Dotsenko et al. \cite{dotsenko2}, who
have shown that the replica symmetry, assumed in the previous application of the method to the RG theory, can be spontaneously broken.
(This kind of replica symmetry breaking (RSB) was previously discovered by Parisi in the theory of spin glasses \cite{parisi}.)

The aim of the present communication is
generalization of the well known  RG equations for the two anisotropic models mentioned above: the model with cubic anisotropy and that with random direction of the anisotropy axis for the case of RSB. We also hope that the way we arrive to such generalization will be of some methodic interest.

\section{RG equations for the case of multicomponent order parameter}

The scalar $\phi^4$ model is described by  the Hamiltonian
\begin{eqnarray}
\label{hamilton}
H[\phi]=\frac{1}{2}\left(\nabla\phi\right)^2+\tau\phi^2 +g\phi^4=H_0+V,
\end{eqnarray}
where $H_0$ represents the first term of the Hamiltonian and $V$ the two other terms; the space has dimension $d$.

We implement the RG by changing the microscopic cut-off implicit in the problem from $a_c$ to $(1+d \ell)a_c$ and asking how the parameters of the model
should be changed to preserve the partition function $Z$.
Considering $V$ as a perturbation one obtains the RG equations for the parameters  \cite{patashinskii}, with the beta functions being  infinite series in the powers of $\tau$ and $g$.
The problem of calculation of the coefficients in the RG equations has two parts: calculation of momentum (coordinate) integrals and calculation of combinatoric multipliers.

The integration part of the problem is exactly the same for  any Hamiltonian with the multicomponent field $\phi_{\cal{A}}$ (calligraphic letter can stand for index of any origin) and quartic interaction.
\begin{eqnarray}
H[\phi]=\frac{1}{2}\sum_{\cal{A}}\left(\nabla\phi_{\cal A}\right)^2 +V(\phi_{\cal A}),
\end{eqnarray}
where $V$ is a polynomial function of  $\phi_{\cal A}$ containing terms of the second and forth degree.
 Only combinatoric multipliers of the perturbation series terms depend upon the specific Hamiltonian.

More specifically, each term in the perturbation series expansion of the beta functions for the generic Hamiltonian is
 the product of that for the scalar $\phi^4$ model, given by integration demanded by the relevant diagram(s),
and
the combinatoric multiplier given by contraction of the interaction (\ref{V}) demanded by this same diagram(s).

To express this idea we can decompose the fields as
\begin{eqnarray}
\phi_{\cal{A}}({\bf r})=\phi({\bf r})\hat{\phi}_{\cal{A}},
\end{eqnarray}
where $\phi({\bf r})$ is a fluctuating basic field, and  $\hat{\phi}_{\cal A}$ shows, so to say, "direction" of the field.
To solve the combinatoric part of the problem we can deal with the symbolic Hamiltonian containing only $\hat{\phi}_{\cal A}$. For example, for the vector
$\phi^4$ model we can use
\begin{eqnarray}
\label{V}
\hat{V}=\sum_a\tau\hat{\phi}^2_a+u\sum_{ab}\hat{\phi}_a^2\hat{\phi}_b^2.
\end{eqnarray}

In general case we have to operate with
\begin{eqnarray}
\label{hamilton1}
\hat{V}= \sum_ig_i\Phi_i,
\end{eqnarray}
where the effective vertices $\Phi_i$ are problem specific  homogeneous polynomials of the second or
fourth  degree of $\hat{\phi}_{\cal A}$  fields.

To first order in $\epsilon$ the RG equations are
\begin{eqnarray}
\label{rgu0}
&&\frac{dg}{d\ell} =\epsilon g-36K_4g^2  \\
\label{rgt07}
&&\frac{d\tau}{d\ell}= 2\tau-12 K_4g\tau,
\end{eqnarray}
where $\epsilon=4-d$, and $K_d=S_d/(2\pi)^d$,  $S_d$ is the surface of the sphere of unit radius in $d$ dimensions.
Recalling that the beta functions are due to box and tadpole diagram respectively we understand that  the RG equations for the generic Hamiltonian can be presented as
\begin{eqnarray}
\label{rgu01}
&&\frac{dg_i}{d\ell} =(d-x_i) g_i-\frac{1}{2}K_4\sum_{jk}c_{ijk}g_jg_k,
\end{eqnarray}
where $x_i$ is the natural scale dimension of $\Phi_i$: $x_i=2d-4$ when $\Phi_i$ is the polynomial of the forth degree, and  $x_i=d-2$ when $\Phi_i$ is the polynomial of the second degree, and
the coefficients $c_{ijk}$  are found from the equation
\begin{eqnarray}
\label{d}
\Phi_j\times \Phi_k=\sum_ic_{ijk}\Phi_i,
\end{eqnarray}
where the sign $\times$ means multiplication and double contraction. (This means that at least one of the effective vertices in the r.h.s. of Eq. (\ref{d}) should be a fourth power polynomial.)

\section{Quenched scalar disorder and cubic anisotropy}
\label{quenched}

Consider $O(n)$ model (the $n$-component order parameter $\phi_i({\bf r})$ $(i=1,2\dots,n)$)
with added  cubic anisotropy  and  quenched scalar disorder terms.
The system is described by the following  Hamiltonian  \cite{ma,cardy,dotsenko}:
\begin{eqnarray}
\label{hamiltonian}
&&H[\delta\tau,\phi]=\frac{1}{2}\sum_{a=1}^n\left(\nabla\phi_a({\bf r})\right)^2
+[\tau-\delta\tau_c({\bf r})]\sum_{a=1}^n\phi_a^2({\bf r})\nonumber\\
&&+u\sum_{a,b=1}^n\phi_a^2({\bf r})\phi_b^2({\bf r})
+v\sum_{a=1}^n\phi_a^4({\bf r}),
\end{eqnarray}
where $\delta\tau$ is a
Gaussian variable of zero mean and variance $\Delta$:
\begin{eqnarray}
<\tau({\bf r})\tau({\bf r}')>=\Delta\delta({\bf r}-{\bf r}').
\end{eqnarray}
 (It is known that fluctuations of the other two coefficients in the Landau-Ginsburg functional do not influence critical behavior for small $\epsilon$. \cite{ma})

\subsection{Replica method}

Replica method is based on the identity
\begin{eqnarray}
F=-\overline{\ln Z}=-\lim_{p\to 0}\frac{\overline{Z^p}-1}{p}\equiv-\lim_{p\to 0}\frac{Z_p-1}{p}.
\end{eqnarray}
Thus,
according to this method  one has to calculate the following partition function (fluctuations of the effective transition temperature $\delta\tau_c({\bf r})$ we assume  to be  Gaussian)
\begin{eqnarray}
&&Z_p=\overline{\left(\int \prod_{a=1}^nD\phi_a\exp\{-\int d^D{\bf r}H[\delta\tau,\phi]\}\right)^p}\nonumber\\
&&=\int D\delta\tau_c({\bf r})\int \prod_{a=1}^n\prod_{\alpha=1}^pD\phi_a^{\alpha}\\
&&\exp\left\{-\int d^d{\bf r} \left[\frac{(\delta\tau_c)^2}{4\Delta}+H[\delta\tau,\phi]\right]\right\},\nonumber
\end{eqnarray}
where the superscript $\alpha$ labels the replicas.

The scheme of the replica method can be described in the following steps \cite{dotsenko}. First, the measurable quantities we are interested in should be calculated for integer $p$.
Second, the analytic continuation of the obtained functions of the parameter $p$ should be made for an arbitrary non-integer $p$. Finally, the limit $p\to 0$ should be taken.

After Gaussian integration over $\delta\tau({\bf r})$ one gets:
\begin{eqnarray}
Z_p=\int \prod_{\cal{A}} D\phi^{\cal{A}}\exp\left\{-\int d^d{\bf r}H[\phi]\right\},
\end{eqnarray}
where
\begin{eqnarray}
\label{z}
&&H=\frac{1}{2}\sum\left(\nabla\phi_a^{\alpha}\right)^2+\tau\sum\left( \phi_a^{\alpha}\right)^2\nonumber\\
&&+\sum \left(\phi_a^{\alpha}\right)^2[(u+v\delta_{ab})\delta_{\alpha\beta}-\Delta]\left(\phi_b^{\beta}\right)^2.
\end{eqnarray}
This time  calligraphic index from Eq. (\ref{hamilton1}) denotes a pair of vector and replica index.

Thus we can deal with
\begin{eqnarray}
\label{hamt94}
\hat{V}=\tau \Phi^1 +u\overline{\Phi}-\Delta \Phi+v\Phi^c,
\end{eqnarray}
where we have introduced four effective vertices
\begin{eqnarray}
\Phi^1&=&\sum_{a\alpha}\left(\hat{\phi}^{\alpha}_a\right)^2\\
\label{Phi}
\Phi&=&\sum_{ab\alpha \beta}\left(\hat{\phi}^{\alpha}_a\right)^2\left(\hat{\phi}^{\beta}_b\right)^2\\
\label{bar}
\overline{\Phi}&=&\sum_{ab\alpha}\left(\hat{\phi}^{\alpha}_a\right)^2\left(\hat{\phi}^{\alpha}_b\right)^2\\
\label{hat}
\Phi^{c}&=&\sum_{a\alpha}\left(\hat{\phi}^{\alpha}_a\right)^4.
\end{eqnarray}

\subsection{RG equations}

In the lowest order in $\epsilon$ we need only  $c_{ijk}$ expansion coefficients.
Substituting the  coefficients from  our multiplication table in the Appendix into Eq. (\ref{rgu01})  we obtain the RG equations
\begin{eqnarray}
\label{c1}
\frac{du}{d\ell}&=&\epsilon u -4K_4\left[(8+n)u^2-12u\Delta+6u v\right]\\
\label{c2}
\frac{d\Delta}{d\ell}&=& \epsilon \Delta-4K_4\left[(4+2n)u\Delta -(8+pn)\Delta^2+6\Delta v\right] \nonumber \\ \\
\label{c3}
\frac{dv}{d\ell}&=& \epsilon v-4K_4[12u v-12 \Delta v+9v^2]
\end{eqnarray}
and
\begin{eqnarray}
\label{ttau5}
\frac{d\tau}{d\ell}=2\tau-4K_4\left[(2+n) u-(2+pn)\Delta+3v\right]\tau.
\end{eqnarray}
The  RG equations coincide with those   of Ref. \onlinecite{aharony2} (for generalization including $\epsilon^2$ terms  see Refs. \onlinecite{lawrie} and  \onlinecite{sarkar}).

The RG equations for the random anisotropy axis model obtained in the framework of the same approach are presented in the Appendix.

\subsection{Replica symmetry breaking}
\label{rsb}

In this Section we consider non-trivial spin-glass effects produced by weak quenched disorder, which have been ignored
in the previous Sections. It is known, that these effects can dramatically change the whole physical scenario for isotropic magnets \cite{dotsenko,dotsenko2}. One can assume that the same can happen for anisotropic magnets.

Consider the ground state properties of the system described by the Hamiltonian (\ref{hamiltonian}). Configurations of the field $\phi$ which
correspond to local minima in $H$ satisfy the saddle-point equation
\begin{eqnarray}
\label{saddle}
-\nabla^2\phi_a+2[\tau-\delta\tau_c({\bf r})]\phi_a+4u\phi_a\sum_b\phi_b^2+4v\phi_a^3=0.\nonumber\\
\end{eqnarray}
The localized solutions of Eq. (\ref{saddle})  with non-zero values of $\phi$ exist in regions of space where $[\tau-\delta\tau_c({\bf r})]$ has negative values. Moreover, for a typical configuration of $\delta\tau_c({\bf r})$, one finds a macroscopic number of local minimum
solutions of the saddle-point equation (\ref{saddle}) \cite{dotsenko}.

In such case one has to take into account the possibility of replica symmetry breaking \cite{dotsenko,dotsenko2}.
We should notice here, that  there exist quite a few studies
where the RSB approach was generalized for situations  with fluctuations (apart from the activities initiated by
Ref. \onlinecite{dotsenko2}, we can cite Refs. \onlinecite{mezard,mezard2,korshunov}), or for the critical phenomena in the Sine-Gordon model  \cite{doussal}.

Detailed analysis of RSB in an $O(n)$ model was performed by Wu \cite{wu}, where
the differential
recursion relations of renormalization group (RG) were derived to the second order of $\epsilon$.
The replicon eigenvalue, which is a simple way to investigate the stability with respect to the
continuous RSB modes, was defined.
It was shown that for $n<n_c=4$ the RG equations do not have  stable fixed point.  For $n>n_c$ only the pure fixed point were found to be physical and stable.

The subject was additionally clarified in the paper by Prudnikov et al. \cite{prudnikov}, where
a field-theoretic description of the critical behavior of the weakly disordered systems is given. Directly, for
three- and two-dimensional systems a renormalization analysis of the effective Hamiltonian of a model with
RSB potentials was carried out in the two-loop approximation.   There a stability of the fixed points
 of the weakly disordered systems with respect to RSB effects  (for a 1-step ansatz) was performed.

A field-theory approach was used to investigate the spin-glass effects on the
critical behaviour of systems with weak temperature-like quenched disorder by Fedorenko \cite{fedorenko}.
There the RG analysis of the effective Hamiltonian of a
model with RSB potentials of a general type was
carried out in the two-loop approximation. The fixed point  stability,
was explored
in terms of replicon eigenvalues. It was  found that the traditional fixed points, which were
usually considered to describe the disorder-induced universal critical behaviour,
remain stable when the continuous RSB modes are taken into account.

Now let us return to our effective Hamiltonian (\ref{z}). RSB is taken into account by considering $\Delta$   not as a constant but as an arbitrary matric $\Delta_{\alpha\beta}$ \cite{dotsenko,dotsenko2}. In this case we can include the term $u\delta_{\alpha\beta}$ in this matrix and present the perturbation $\hat{V}$ as as
\begin{eqnarray}
\label{z2}
\hat{V}=\tau \Phi^1  -\sum_{\alpha\beta}\Delta_{\alpha\beta} \Phi_{\alpha\beta}+v\Phi^c,
\end{eqnarray}
where we have introduced additional vertex
\begin{eqnarray}
\label{Phir}
\Phi_{\alpha\beta}=\sum_{ab}\left(\hat{\phi}^{\alpha}_a\right)^2\left(\hat{\phi}^{\beta}_b\right)^2.
\end{eqnarray}

We shall write down RG equations assuming that the matrix $\Delta_{\alpha\beta}$
has a general Parisi RSB structure, and in the limit $p\to 0$ is parameterized in terms of its
diagonal elements $\widetilde{\Delta}$ and the off-diagonal function $\Delta(x)$ defined in the interval $0<x<0$ (which can be presented as $\Delta=(\widetilde{\Delta},\Delta(x)$). \cite{dotsenko}.

The standard technique of the Parisi RSB algebra is  different from ordinary matrix algebra in defining the  product of matrices  \cite{parisi}
(and we have such product in the r.h.s. of Eq. (\ref{c2})). The  definition of the product of  Parisi matrices  is as follows.
Let $a=(\widetilde{a},a(x))$, $b=(\widetilde{b},b(x))$, $c=(\widetilde{c},c(x))$, and $c=ab$. Then
\begin{eqnarray}
\label{hr}
&&\widetilde{c}=\widetilde{a}\widetilde{b}-\int_0^1dxa(x)b(x)\\
\label{en}
&&c(x)=\left(\widetilde{a}-\int_0^1dya(y)\right)b(x)+\left(\widetilde{b}-\int_0^1dyb(y)\right)a(x)\nonumber\\
&&-\int_0^xdy[a(x)-a(y)][b(x)-b(y)].
\end{eqnarray}

Thus the RG equations  become:
\begin{eqnarray}
\label{c1c}
\frac{d\widetilde{\Delta}}{d\ell}&=& \epsilon \widetilde{\Delta}+4K_4\left[(8+n)\widetilde{\Delta}^2-n\int_0^1dx\Delta^2(x)-6v\widetilde{\Delta}\right] \nonumber\\ \\
\label{c2c}
\frac{d\Delta(x)}{d\ell}&=& \epsilon \Delta(x)+4K_4\left[(4+2n)\widetilde{\Delta}\Delta(x)+4\Delta^2(x)\right.\nonumber\\
&-&2n\Delta(x)\int_0^1dy\Delta(y)-n\int_0^xdy[\Delta(x)-\Delta(y)]^2\nonumber\\
&-&\left. 6v\Delta(x)\right]  \\
\label{c3c}
\frac{dv}{d\ell}&=& \epsilon v+4K_4[12 v\widetilde{\Delta}-9v^2].
\end{eqnarray}

Eqs. (\ref{c1c})-(\ref{c3c}) obviously have replica symmetric fixed points, which are the solutions of the system
\begin{eqnarray}
\label{c1cb}
&& \epsilon \widetilde{\Delta}_0+4K_4\left[(8+n)\widetilde{\Delta}_0^2-n\Delta_0^2-6v_0\widetilde{\Delta}_0\right]=0 \\
\label{c2cb}
&&\epsilon \Delta_0+4K_4\left[(4+2n)\widetilde{\Delta}_0\Delta_0+(4-2n)\Delta_0^2- 6v_0\Delta_0\right]=0 \nonumber\\  \\
\label{c3cb}
&&\epsilon v_0+4K_4[12 v_0\widetilde{\Delta}_0-9v_0^2]=0.
\end{eqnarray}
Stability of these fixed point with respect to replica symmetric deviations is a well studied problem \cite{aharony}.   
The aim of the present paper is to study stability of the fixed points with respect to small deviations which may not be (or, in a particular case, may be) replica symmetric. Assuming 
\begin{eqnarray}
\widetilde{\Delta}&=&\widetilde{\Delta}_0+\delta\widetilde{\Delta}\\
v&=&v_0+\delta v\\
\Delta(x)&=&\Delta_0+\delta\Delta(x)
\end{eqnarray}
and linearizing Eqs. (\ref{c1c})-(\ref{c3c}) with respect to small deviations we obtain
\begin{eqnarray}
\label{c1cd}
&&\frac{d\delta\widetilde{\Delta}}{d\ell}= \epsilon \delta\widetilde{\Delta}+8K_4\left[(8+n)\widetilde{\Delta}_0\delta\widetilde{\Delta}\right.\nonumber\\
&&-\left. n\Delta_0\int_0^1dx\delta\Delta(x)-3v_0\delta\widetilde{\Delta}-3\widetilde{\Delta}_0\delta v\right]  \\
\label{c2cd}
&&\frac{d\delta\Delta(x)}{d\ell}= \epsilon \delta\Delta(x)+4K_4\left[(4+2n)\widetilde{\Delta}_0\delta\Delta(x)\right.\nonumber\\
&&+(4+2n)\Delta_0\delta\widetilde{\Delta}
+8\Delta_0\delta\Delta(x)-2n\Delta_0\int_0^1dy\delta\Delta(y)\nonumber\\
&&\left.-2n\Delta_0\delta\Delta(x)- 6v_0\delta\Delta(x)- 6\Delta_0\delta v\right]  \\
\label{c3cd}
&&\frac{d\delta v}{d\ell}= \epsilon \delta v+8K_4[6 v_0\delta\widetilde{\Delta}+6 \widetilde{\Delta}_0\delta v-9v_0\delta v].
\end{eqnarray}
Integrating Eq. (\ref{c2cd}) we obtain
\begin{eqnarray}
\label{c2ce}
&&\frac{d\delta\Delta}{d\ell}= \epsilon \delta\Delta+4K_4\left[(4+2n)\widetilde{\Delta}_0\delta\Delta\right.\nonumber\\
&&+(4+2n)\Delta_0\delta\widetilde{\Delta}
+8\Delta_0\delta\Delta-2n\Delta_0\delta\Delta\nonumber\\
&&\left.-2n\Delta_0\delta\Delta- 6v_0\delta\Delta- 6\Delta_0\delta v\right],
\end{eqnarray}
where
\begin{eqnarray}
\delta\Delta=\int_0^1dx\delta\Delta(x).
\end{eqnarray}
Eqs. (\ref{c1cd}), (\ref{c2ce}), (\ref{c3cd}) tell us that the stability of a fixed point with respect to deviations without replica symmetry is the same as the stability of the fixed point with respect to replica symmetric deviations.

\section{Discussion}

The main result of this paper is the RG equations taking into account possible RSB for the model with cubic anisotropy and quenched scalar disorder
 (Eqs. (\ref{c1c}) - (\ref{c3c})). They present the generalization of those obtained
for the case of ferromagnetic transition in vector $\phi^4$ model with quenched scalar disorder (no cubic anisotropy term) \cite{dotsenko, dotsenko2}. 
We study stability of the replica symmetric fixed point of the RG equations with respect to the deviations without replica symmetry.
We find that the stability is the same in both cases, which may be considered as the generalization of the conditions of applicability of the canonical results. Whether there exist fixed points of the RG equations without replica symmetry remains an open question.

\begin{acknowledgments}

We see our modest contribution as one more illustration to the famous saying of Leopold Kronecker:
  `Die ganzen Zahlen hat der liebe Gott gemacht, alles andere ist Menschenwerk' (`God made the integers, all else is the work of man').

One of the authors (E.K.) cordially thanks  for the hospitality extended to him during
his stay: Max-Planck-Institut fur Physik komplexer Systeme, where the work was initiated, and
Center for Theoretical Physics of Complex Systems, where the work continued.

Discussions with A. Aharony, J. Cardy, J. Holland, I. D. Lawrie, F. Pollmann, N. Sarkar,  A. Sinner, and K. Ziegler   are gratefully acknowledged.
\end{acknowledgments}

\appendix

\section{Vertices contractions}

We have seen that the problem of deriving the RG equations is reduced to combinatoric problem, the latter consists in contraction the vortices.
Explicitly performing such contractions   we obtain the multiplication table used in Section \ref{quenched}
\begin{eqnarray}
\label{first}
\Phi^1\times\Phi &=& 4(2+pn)\Phi^1    \\
\Phi^1\times\overline{\Phi} &=&4(2+n)\Phi^1\\
\Phi^1\times\Phi^c&=& 12\Phi^1\\
\label{phii}
\Phi\times \Phi&=&8(8+pn)\Phi\\
\Phi\times\overline{\Phi} &=&8(2+n)\Phi+48\overline{\Phi}\\
\Phi\times\Phi^c&=& 24\Phi+48\Phi^c\\
\overline{\Phi}\times\overline{\Phi} &=&8(8+n)\overline{\Phi}\\
\overline{\Phi}\times\Phi^c&=&24\overline{\Phi}+48\Phi^c \\
\label{last}
\Phi^c\times\Phi^c&=&72\Phi^c.
\end{eqnarray}

In Section \ref{simple} we used additional identities:
\begin{eqnarray}
\label{ph}
\Phi^1\times \Psi&=&4(1+p+n)\Phi^1\\
\Phi\times \Psi&=&8(1+p+n)\Phi+48\Psi\\
\label{phipsi}
\overline{\Phi}\times \Psi&=&8\Phi+8(5+n)\overline{\Phi}+16\Psi\\
\label{phipsi6}
\Psi\times \Psi&=&8(4+p+n)\Psi+24\Phi.
\end{eqnarray}
The derivation can be  graphically presented as

\setlength{\unitlength}{.6cm}
\begin{picture}(6,7)
\thicklines



\put(0,4.7){$\Phi^1\times\Psi=4$}
\put(4.3,4.9){\oval(2,1.8)}
\put(6.3,4.9){\oval(2,1.8)[l]}
\put(3.3,4.9){\circle*{.4}}
\put(5.3,4.9){\circle*{.4}}
\put(4,3.5){$a\alpha$}
\put(4,6){$a\alpha$}
\put(5.7,3.5){$a\alpha$}
\put(5.7,6){$a\alpha$}
\put(3.7,4.7){$\Phi^1$}
\put(5.7,4.7){$\Psi$}


\put(7,4.7){$+4$}
\put(11.3,4.9){\oval(2,1.8)[l]}
\put(9.3,4.9){\oval(2,1.8)}
\put(8.3,4.9){\circle*{.4}}
\put(10.3,4.9){\circle*{.4}}
\put(9,3.5){$a\alpha$}
\put(9,6){$a\alpha$}
\put(10.7,3.5){$a\beta$}
\put(10.7,6){$a\beta$}
\put(8.7,4.7){$\Phi^1$}
\put(10.7,4.7){$\Psi$}


\put(2,1.2){$+4$}
\put(4.3,1.4){\oval(2,1.8)}
\put(6.3,1.4){\oval(2,1.8)[l]}
\put(3.3,1.4){\circle*{.4}}
\put(5.3,1.4){\circle*{.4}}
\put(4,0){$a\alpha$}
\put(4,2.5){$a\alpha$}
\put(5.7,0){$b\alpha$}
\put(5.7,2.5){$b\alpha$}
\put(3.7,1.2){$\Phi^1$}
\put(5.7,1.2){$\Psi$}
\put(7,1.2){$=4(1+p+n)\Phi^1$}

\end{picture}

\setlength{\unitlength}{.6cm}
\begin{picture}(6,7)
\thicklines



\put(0.2,4.7){$\Phi\times\Psi=8$}
\put(2.8,4.9){\oval(2,1.8)[r]}
\put(4.8,4.9){\oval(2,1.8)}
\put(6.8,4.9){\oval(2,1.8)[l]}
\put(3.8,4.9){\circle*{.4}}
\put(5.8,4.9){\circle*{.4}}
\put(2.8,3.5){$a\alpha$}
\put(2.8,6){$a\alpha$}
\put(4.5,3.5){$b\beta$}
\put(4.5,6){$b\beta$}
\put(6.2,3.5){$b\beta$}
\put(6.2,6){$b\beta$}
\put(4.2,4.7){$\Phi$}
\put(6.2,4.7){$\Psi$}


\put(7.2,4.7){$+8$}
\put(8,4.9){\oval(2,1.8)[r]}
\put(10,4.9){\oval(2,1.8)}
\put(12,4.9){\oval(2,1.8)[l]}
\put(9,4.9){\circle*{.4}}
\put(11,4.9){\circle*{.4}}
\put(8,3.5){$a\alpha$}
\put(8,6){$a\alpha$}
\put(9.7,3.5){$b\beta$}
\put(9.7,6){$b\beta$}
\put(11.4,3.5){$b\gamma$}
\put(11.4,6){$b\gamma$}
\put(9.4,4.7){$\Phi$}
\put(11.4,4.7){$\Psi$}



\put(2,1.2){$+8$}
\put(2.8,1.4){\oval(2,1.8)[r]}
\put(4.8,1.4){\oval(2,1.8)}
\put(6.8,1.4){\oval(2,1.8)[l]}
\put(3.8,1.4){\circle*{.4}}
\put(5.8,1.4){\circle*{.4}}
\put(2.8,0){$a\alpha$}
\put(2.8,2.5){$a\alpha$}
\put(4.5,0){$b\beta$}
\put(4.5,2.5){$b\beta$}
\put(6.2,0){$c\beta$}
\put(6.2,2.5){$c\beta$}
\put(4.2,1.2){$\Phi$}
\put(6.2,1.2){$\Psi$}


\put(7.2,1.2){$+48$}
\put(8,1.4){\oval(2,1.8)[r]}
\put(10,1.4){\oval(2,1.8)}
\put(12,1.4){\oval(2,1.8)[l]}
\put(9,1.4){\circle*{.4}}
\put(11,1.4){\circle*{.4}}
\put(8,0){$a\alpha$}
\put(8,2.5){$b\beta$}
\put(9.7,0){$a\alpha$}
\put(9.7,2.5){$b\beta$}
\put(11.4,0){$b\alpha$}
\put(11.4,2.5){$a\beta$}
\put(9.4,1.2){$\Phi$}
\put(11.4,1.2){$\Psi$}

\end{picture}

\setlength{\unitlength}{.6cm}
\begin{picture}(6,2)
\thicklines

\put(2.2,.8){$=8(1+p+n)\Phi+48$}

\end{picture}

\setlength{\unitlength}{.6cm}
\begin{picture}(6,7)
\thicklines



\put(0.2,4.7){$\overline{\Phi}\times \Psi=8$}
\put(2.8,4.9){\oval(2,1.8)[r]}
\put(4.8,4.9){\oval(2,1.8)}
\put(6.8,4.9){\oval(2,1.8)[l]}
\put(3.8,4.9){\circle*{.4}}
\put(5.8,4.9){\circle*{.4}}
\put(2.8,3.5){$a\alpha$}
\put(2.8,6){$a\alpha$}
\put(4.5,3.5){$b\alpha$}
\put(4.5,6){$b\alpha$}
\put(6.2,3.5){$b\alpha$}
\put(6.2,6){$b\alpha$}
\put(4.2,4.7){$\overline{\Phi}$}
\put(6.2,4.7){$\Psi$}


\put(7.2,4.7){$+8$}
\put(8,4.9){\oval(2,1.8)[r]}
\put(10,4.9){\oval(2,1.8)}
\put(12,4.9){\oval(2,1.8)[l]}
\put(9,4.9){\circle*{.4}}
\put(11,4.9){\circle*{.4}}
\put(8,3.5){$a\alpha$}
\put(8,6){$a\alpha$}
\put(9.7,3.5){$b\alpha$}
\put(9.7,6){$b\alpha$}
\put(11.4,3.5){$b\beta$}
\put(11.4,6){$b\beta$}
\put(9.4,4.7){$\overline{\Phi}$}
\put(11.4,4.7){$\Psi$}



\put(2,1.2){$+8$}
\put(2.8,1.4){\oval(2,1.8)[r]}
\put(4.8,1.4){\oval(2,1.8)}
\put(6.8,1.4){\oval(2,1.8)[l]}
\put(3.8,1.4){\circle*{.4}}
\put(5.8,1.4){\circle*{.4}}
\put(2.8,0){$a\alpha$}
\put(2.8,2.5){$a\alpha$}
\put(4.5,0){$b\alpha$}
\put(4.5,2.5){$b\alpha$}
\put(6.2,0){$c\alpha$}
\put(6.2,2.5){$c\alpha$}
\put(4.2,1.2){$\overline{\Phi}$}
\put(6.2,1.2){$\Psi$}


\put(7.2,1.2){$+16$}
\put(8,1.4){\oval(2,1.8)[r]}
\put(10,1.4){\oval(2,1.8)}
\put(12,1.4){\oval(2,1.8)[l]}
\put(9,1.4){\circle*{.4}}
\put(11,1.4){\circle*{.4}}
\put(8,0){$a\alpha$}
\put(8,2.5){$b\alpha$}
\put(9.7,0){$a\alpha$}
\put(9.7,2.5){$b\alpha$}
\put(11.4,0){$a\beta$}
\put(11.4,2.5){$b\beta$}
\put(9.4,1.2){$\overline{\Phi}$}
\put(11.4,1.2){$\Psi$}

\end{picture}

\setlength{\unitlength}{.6cm}
\begin{picture}(6,3.3)
\thicklines



\put(2,1.2){$+32$}
\put(2.8,1.4){\oval(2,1.8)[r]}
\put(4.8,1.4){\oval(2,1.8)}
\put(6.8,1.4){\oval(2,1.8)[l]}
\put(3.8,1.4){\circle*{.4}}
\put(5.8,1.4){\circle*{.4}}
\put(2.8,0){$a\alpha$}
\put(2.8,2.5){$a\alpha$}
\put(4.5,0){$a\alpha$}
\put(4.5,2.5){$b\alpha$}
\put(6.2,0){$a\alpha$}
\put(6.2,2.5){$b\alpha$}
\put(4.2,1.2){$\overline{\Phi}$}
\put(6.2,1.2){$\Psi$}


\put(7.2,1.2){$=8\Phi+8(5+n)\overline{\Phi}+16\Psi$}

\end{picture}

\setlength{\unitlength}{.6cm}
\begin{picture}(6,7)
\thicklines



\put(0.2,4.7){$\Psi\times\Psi=8$}
\put(2.8,4.9){\oval(2,1.8)[r]}
\put(4.8,4.9){\oval(2,1.8)}
\put(6.8,4.9){\oval(2,1.8)[l]}
\put(3.8,4.9){\circle*{.4}}
\put(5.8,4.9){\circle*{.4}}
\put(2.8,3.5){$a\alpha$}
\put(2.8,6){$b\alpha$}
\put(4.5,3.5){$a\beta$}
\put(4.5,6){$b\beta$}
\put(6.2,3.5){$a\gamma$}
\put(6.2,6){$b\gamma$}
\put(4.2,4.7){$\Psi$}
\put(6.2,4.7){$\Psi$}


\put(7.2,4.7){$+8$}
\put(8,4.9){\oval(2,1.8)[r]}
\put(10,4.9){\oval(2,1.8)}
\put(12,4.9){\oval(2,1.8)[l]}
\put(9,4.9){\circle*{.4}}
\put(11,4.9){\circle*{.4}}
\put(8,3.5){$a\alpha$}
\put(8,6){$a\beta$}
\put(9.7,3.5){$b\alpha$}
\put(9.7,6){$b\beta$}
\put(11.4,3.5){$c\alpha$}
\put(11.4,6){$c\beta$}
\put(9.4,4.7){$\Psi$}
\put(11.4,4.7){$\Psi$}



\put(2,1.2){$+16$}
\put(2.8,1.4){\oval(2,1.8)[r]}
\put(4.8,1.4){\oval(2,1.8)}
\put(6.8,1.4){\oval(2,1.8)[l]}
\put(3.8,1.4){\circle*{.4}}
\put(5.8,1.4){\circle*{.4}}
\put(2.8,0){$a\alpha$}
\put(2.8,2.5){$b\alpha$}
\put(4.5,0){$a\beta$}
\put(4.5,2.5){$b\beta$}
\put(6.2,0){$a\beta$}
\put(6.2,2.5){$b\beta$}
\put(4.2,1.2){$\Psi$}
\put(6.2,1.2){$\Psi$}


\put(7.2,1.2){$+16$}
\put(8,1.4){\oval(2,1.8)[r]}
\put(10,1.4){\oval(2,1.8)}
\put(12,1.4){\oval(2,1.8)[l]}
\put(9,1.4){\circle*{.4}}
\put(11,1.4){\circle*{.4}}
\put(8,0){$a\alpha$}
\put(8,2.5){$b\beta$}
\put(9.7,0){$b\alpha$}
\put(9.7,2.5){$b\beta$}
\put(11.4,0){$a\beta$}
\put(11.4,2.5){$b\beta$}
\put(9.4,1.2){$\Psi$}
\put(11.4,1.2){$\Psi$}

\end{picture}

\setlength{\unitlength}{.6cm}
\begin{picture}(6,3.5)
\thicklines




\put(2,1.2){$+24$}
\put(2.8,1.4){\oval(2,1.8)[r]}
\put(4.8,1.4){\oval(2,1.8)}
\put(6.8,1.4){\oval(2,1.8)[l]}
\put(3.8,1.4){\circle*{.4}}
\put(5.8,1.4){\circle*{.4}}
\put(2.8,0){$a\alpha$}
\put(2.8,2.5){$b\beta$}
\put(4.5,0){$b\alpha$}
\put(4.5,2.5){$a\beta$}
\put(6.2,0){$a\alpha$}
\put(6.2,2.5){$b\beta$}
\put(4.2,1.2){$\Psi$}
\put(6.2,1.2){$\Psi$}


\put(7.2,1.2){$=8(4+p+n)\Psi+24\Phi$}

\end{picture}

\section{RG equations to $\epsilon^2$ order}

RG equations for the scalar $\phi^4$ model to $\epsilon^2$ order are:
\begin{eqnarray}
&&\frac{dg}{d\ell} =\epsilon g-36K_4g^2+816K_4^2g^3  \\
&&\frac{d\tau}{d\ell}= 2\tau-12 K_4g\tau+120K_4^2g^2\tau \\
\label{zz}
&&\frac{dZ}{d\ell} =-12K_4^2g^2Z;
\end{eqnarray}
Eq. (\ref{zz}) describes renormalization of the fields. For the generic Hamiltonian (\ref{hamilton1}) we have
\begin{eqnarray}
\label{rgu}
&&\frac{dg_i}{d\ell} =(d-x_i) g_i-\frac{1}{2}K_4\sum_{jk}c_{ijk}g_jg_k \nonumber\\
&&+\frac{1}{3!}K_4^2\sum_{jkl}c_{ijkl}g_jg_kg_l-\frac{1}{8}K_4^2g_i\sum_{{\cal A}jk}d_{{\cal A}jk}g_jg_k\\
\label{zz0}
&&\frac{dZ_{\cal A}}{d\ell} =-\frac{1}{16}K_4^2\sum_{ij}d_{{\cal A}ij}g_ig_jZ_{\cal A}.
\end{eqnarray}
The coefficients $d_{{\cal A}ij}$ are found  from the equation
\begin{eqnarray}
\label{d5h}
\Phi_i\times \Phi_j=\sum_{\cal A}d_{{\cal A}ij}\phi_{\cal A}^2;
\end{eqnarray}
the sign $\times$ in Eq. (\ref{d5h}) means multiplication and contraction three times. (This means that both effective vertices in the r.h.s. of Eq. (\ref{d5h}) should be fourth power polynomials.)
The coefficients  $c_{ijkl}$ are found  from the equation
\begin{eqnarray}
\label{d5h2}
\Phi_j\times \Phi_k\times\Phi_l=\sum_ic_{ijkl}\Phi_i;
\end{eqnarray}
in Eq. (\ref{d5h2}) contraction is performed four times. Summation with respect to ${\cal A}$ in the last term in the r.h.s. of Eq. (\ref{rgu}) is performed with respect to all fields entering into $\Phi_i$.

\section{Phase portraits}

\begin{figure}[h]
\includegraphics[width= .7\columnwidth, clip=true]{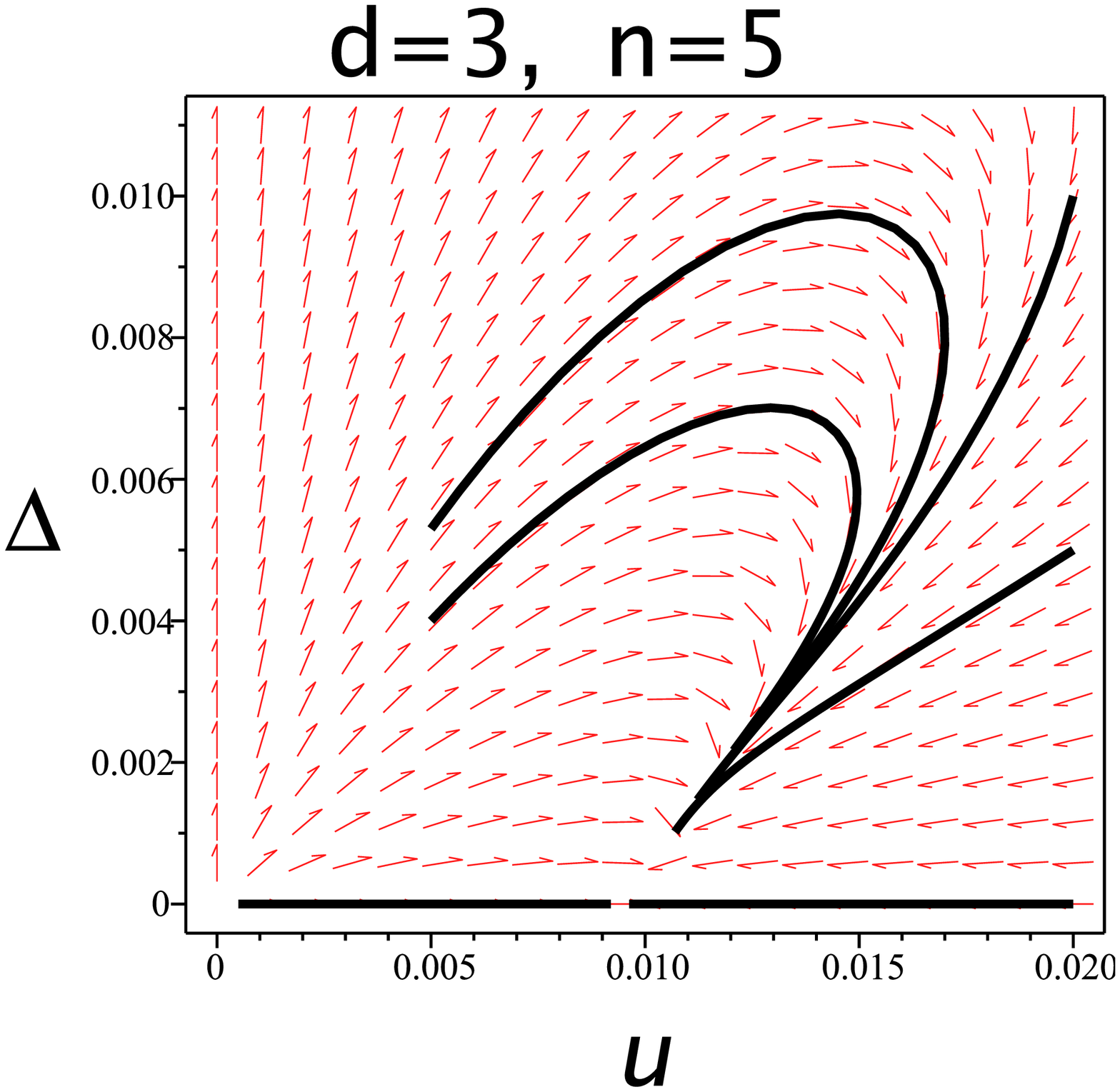}
\includegraphics[width= .7\columnwidth, clip=true]{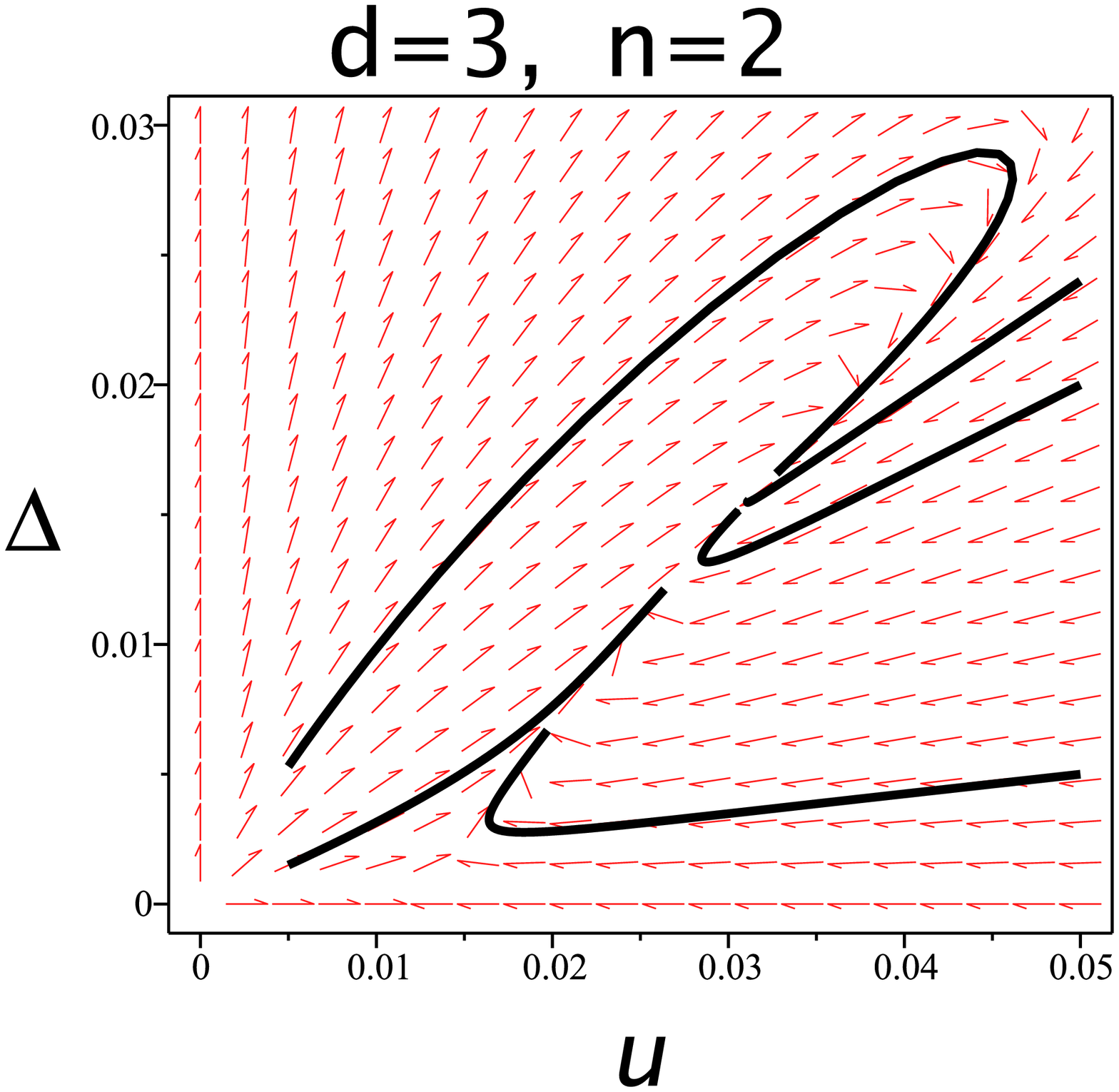}
\caption{\label{fig:d3n2} Phase portraits of the replica symmetric RG equations for the $n$-component vector  $\phi^4$ model in $d$ dimensions with quenched disorder (Eqs. (\ref{c1}),(\ref{c2}) with $p=0$ and $v=0$). }
\end{figure}

As a simple illustration we present on   Fig. \ref{fig:d3n2} phase  portraits of the system  (\ref{c1}),(\ref{c2})  without cubic anisotropy  ($v=0$) in RS subspace.  We see the stable    pure  fixed points for  $n=5$ and the stable  random  fixed points for  $n=2$. In both cases we see  the  unstable Gaussian fixed point.

\section{Random direction of the anisotropy axis}
\label{simple}

The random-axis model
\begin{eqnarray}
\label{anisotropy}
H=-J\sum_{ij}J_{ij}\vec{S}_i\vec{S}_j-D_0\sum_i(\vec{x}_i\vec{S}_i)^2
\end{eqnarray}
was introduced by Harris et al. \cite{harris}
 to describe the magnetic
properties of amorphous alloys. In Eq. (\ref{anisotropy}) $\vec{S}_i$ is an $n$-component spin vector located at the lattice site $i$, $J_{ij}$ is the exchange interaction,  $\hat{x}_i$ is a unit vector which points in the local
(random) direction of the uniaxial anisotropy at the
site $i$, and $D_0$ is the anisotropy constant.

The Hamiltonian of the model in the continuum approximation and after the replica trick can be presented as \cite{aharony2}
\begin{eqnarray}
\label{hamt9}
&&H=\frac{1}{2}\sum\left(\nabla\phi_a^{\alpha}\right)^2+\tau\sum\left( \phi_a^{\alpha}\right)^2\\
&&+\sum \left[\left(\phi_a^{\alpha}\right)^2(u\delta_{\alpha\beta}-\Delta)\left(\phi_b^{\beta}\right)^2
+w\phi^{\alpha}_a\phi^{\alpha}_b\phi^{\beta}_a\phi^{\beta}_b\right].\nonumber
\end{eqnarray}
Thus we can deal with
\begin{eqnarray}
\hat{V}=\tau \Phi^1 +u\overline{\Phi}-\Delta \Phi+w\Psi,
\end{eqnarray}
where
\begin{eqnarray}
\label{b3}
\Psi=\sum\hat{\phi}^{\alpha}_a\hat{\phi}^{\alpha}_b\hat{\phi}^{\beta}_a\hat{\phi}^{\beta}_b.
\end{eqnarray}
The connection between the parameters of the Hamiltonians (\ref{anisotropy}) and (\ref{hamt9}) will be of no interest to us.

Elementary algebra gives additional lines of the
multiplication table necessary for obtaining the RG equations in the case considered (see Appendix).
Thus  we obtain the RG equations
\begin{eqnarray}
\label{c1b}
&&\frac{du}{d\ell}=\epsilon u -4K_4\left[(8+n)u^2-12u\Delta +2(5+n)uw\right]\nonumber\\\\
\label{c2b}
&&\frac{d\Delta}{d\ell}= \epsilon \Delta-4K_4\left[(4+2n)u\Delta -(8+pn)\Delta^2 \right.\nonumber\\
&&\left. -2uw+2(1+p+n)\Delta w-3w^2\right]\\
\label{c3b}
&&\frac{dw}{d\ell}= \epsilon w -4K_4\left[4uw-12\Delta w+(4+p+n)w^2\right].\nonumber\\
\end{eqnarray}
The analog of Eqs. (\ref{ttau5})  is
\begin{eqnarray}
\label{ttau2}
\frac{d\tau}{d\ell}=2\tau-4K_4\left[(2+n) u-(2+pn)\Delta+(1+p+n)w\right]\tau.\nonumber\\
\end{eqnarray}
Eqs. (\ref{c1b}) - (\ref{ttau2})  exactly coincide with those from Ref. \cite{aharony2}.
However, these equations do not give physically relevant stable fixed point \cite{aharony2,dudka}.
If no such fixed point exists
one usually concludes that the system does
not show a second order phase transition but a
first order phase transition (this is especially the
case when one finds runaway solutions of
the RG equations; one prominent physical example
being the transition to the superconducting
phase) \cite{dudka}. Alternatively, the low temperature phase can be a spin-glass and not a ferromagnet.

\section{Operator product expansion}

Eq. (\ref{rgu01}) appear naturally within the operator product expansion (OPE) method, another great discovery of Wilson \cite{wilson2} (see also Refs. \onlinecite{peskin,holland}; application of this method
 to the theory of classical phase transitions is particularly clearly presented  in the  book by Cardy \cite{cardy}).

The operator product expansion is a  universal conception of quantum field theory.
The essential idea is that for any two local operator quantum fields at  points ${\bf x},{\bf y}$ (we consider Euclidean space) their product may be
expressed in terms of a series of local quantum fields at any other point  ( which may be identified with ${\bf x}$ or ${\bf y}$)
times $c$-number coefficient functions which depend on $|{\bf x}-{\bf y}|$.

This general statement, in particular case that will be relevant for us, can be presented as follows \cite{patashinskii,cardy}.
Let  $\Phi$ (called scaling field)  be some product of massless free fields.
 Then
\begin{eqnarray}
\label{rcl}
&&  :\Phi_i({\bf x}):\times:\Phi_j({\bf y}):
  =  :\Phi_i({\bf x}) \Phi_j({\bf x}): +\nonumber \\
&& +  \sum_{1-contraction\ [({\bf x},i)({\bf y},j)]} \Delta_{\cal{AB}}(|{\bf x}-{\bf y}|)  :\Phi_i({\bf x})\Phi_j({\bf x}): \nonumber \\
&& +\sum_{2-contractions\ [({\bf x},i)({\bf y},j)],[({\bf x},i')({\bf y},j')]}\
\Delta_{\cal{AB}} (|{\bf x}-{\bf y}|)\nonumber\\
 &&  \Delta_{\cal{A}'\cal{B}'} (|{\bf x}-{\bf y}|):\Phi_i({\bf x})\Phi_j({\bf x}):+ \dots,
\end{eqnarray}
where
$:X :$  stands for normal ordered operator  $X$, and
\begin{eqnarray}
\Delta_{\cal{AB}}(x)=\frac{\delta_{AB}}{4\pi}\frac{\Gamma(\sigma)}{\pi^{\sigma}}\frac{1}{x^{2\sigma}},\;\;\;\sigma=d/2-1
\end{eqnarray}
 is the propagator of the free fields.(Further on, not to clutter notation, we'll omit the colon signs, where it can not lead to confusion.)

Let us consider a fixed point Hamiltonian $H^*$ which is perturbed by a number of scaling fields, so that the partition function is \cite{cardy}
\begin{eqnarray}
Z=\text{Tr}\; \exp\left\{-\int d^d{\bf r}\left[H^*+\sum_ia_c^{x_i}g_i\Phi_i({\bf r})\right]\right\},
\end{eqnarray}
where $x_i$ is the appropriate natural scaling dimension, and   microscopic cut-off   $a_c$  is  implied in the integral.
Expanding in the powers of coupling we obtain
\begin{eqnarray}
&&Z=Z^*\left[1-\sum_i a_c^{x_i-d}g_i\int d^d{\bf r}\langle\Phi_i({\bf r})\rangle\right.\\
&&\left.+\frac{1}{2}\sum_{ij}a_c^{x_i+x_j-2d}g_ig_j\int d^d{\bf r}_1d^d{\bf r}_2\langle\Phi_i({\bf r}_1)\Phi_j({\bf r}_2)\rangle-\dots\right],\nonumber
\end{eqnarray}
where all correlation functions are to be evaluated with respect to the fixed point Hamiltonian $H^*$.

We implement the RG by changing the microscopic cut-off from $a_c$ to $(1+d \ell)a_c$ and asking how the couplings $g_i$
should be changed to preserve the partition function $Z$. The answer is given by the perturbative RG equations  \cite{cardy}
\begin{eqnarray}
\label{sc}
\frac{dg_k}{d\ell}=(d-x_k)g_k-\frac{1}{2}K_d\sum_{ij}c_{kij}g_ig_j+\dots,
\end{eqnarray}
where summation is with respect to all pairs $i,j$ such, that $\Phi_k$ appears in the product $\Phi_i\times\Phi_j$ as the result of 2 contractions.

\end{document}